\documentclass[twocolumn,english,reprint,prl,aps,superscriptaddress,nofootinbib,floatfix]{revtex4-1}
\usepackage[latin9]{inputenc}
\usepackage{babel}
\usepackage{bm}
\usepackage{amsmath}
\usepackage{amssymb}
\usepackage[unicode=true,]
 {hyperref}

\makeatletter

\DeclareTextSymbolDefault{\textquotedbl}{T1}

\usepackage{graphicx}
\usepackage{bm}\usepackage{color}\usepackage{amsfonts}\usepackage{caption}\usepackage{float}\usepackage{comment}%\usepackage[T1]{fontenc}
\usepackage{wasysym}
\usepackage{babel}
\usepackage{amsfonts}

\makeatother

\begin{document}
\title{Gravitons in a Box}
\author{Sougato Bose}
\affiliation{Department of Physics and Astronomy, University College London, Gower
Street, WC1E 6BT London, UK}
\author{Anupam Mazumdar}
\affiliation{Van Swinderen Institute, University of Groningen, 9747 AG, The Netherlands}
\author{Marko Toro\v{s}}
\affiliation{School of Physics and Astronomy, University of Glasgow, Glasgow, G12
8QQ, UK}
\begin{abstract}
Gravity and matter are universally coupled, and this unique universality
provides us with an intriguing way to quantifying quantum aspects
of space-time in terms of the number of gravitons 
within a given box. In particular, we will provide a limit on the
the number of gravitons if we trace out the matter degrees of freedom.
We will obtain the universal bound on the number of gravitons,
which would be given by $N_{g}\approx(m/M_{p})^{2}$. Since the number
of gravitons also signifies the number of bosonic states they occupy,
the number of gravitons will indirectly constrain the system's gravitational entropy. We will show that it saturates Bekenstein bound on the
gravitational Area-law of entropy. Based on these observations, we will ascertain that the gravitons permeating in the {\it observable} Universe always $N_g\gg 1$. 
\end{abstract}
\maketitle
%%%%%%%%%%%%%%%%%%%%%%%%%%%%%%%%%%%%%%%%

\textbf{Introduction:} Within the general theory of relativity, the
dynamics of space-time is intertwined by both matter and the gravitational
degrees of freedom. If both matter and the gravitational degrees of
freedom are quantum in nature, then their interplay will pave an important
role in any quantum theory of gravity~\cite{kiefer}. However, there
is no experimental proof yet to show that gravity is quantum in nature.

Like other forces of nature, the spin-2 \textit{graviton} is thought
to be the carrier of the gravitational interaction, and it is responsible
for the gravitational attraction between two massive bodies. A graviton
can be canonically quantised around a weak curvature background~\cite{Gupta:1954zz},
and we can attribute both on-shell and off-shell degrees of freedom
to the massless graviton. The former is responsible for describing
independent dynamical modes, while the latter describes
how the force is being mediated between two massive bodies. One of
the interesting features of the gravitational interaction is that
it is universal, and it is governed by Newton's constant $G\sim1/M_{p}^{2}$,
where $M_{p}\sim1.2\times10^{19}$GeV.

Niels Bohr once argued, double-slit experiment with an electron, that
the photon which mediates electromagnetic interaction ought to be
quantum if the electron is a bonafide quantum entity, see~\cite{Baym:2009zu}.
The purported weakness of the universal gravitational interaction
with the matter precluded a similar argument favouring the quantum
nature of a graviton. A similar weakness in the gravitational interaction
was poignantly used by Dyson to point out that it would be extremely
challenging to detect gravitons~\cite{Dyson}. Although, very tiny
but a quantum graviton interaction with matter can leave indelible
mark in classical/quantum systems~\cite{Ng:1999hm,AmelinoCamelia:1997gz,Parikh:2020nrd,Toros:2020dbf,Bose:2017nin,Vedral}.

In fact, just quantum mechanics, along with the special theory of
relativity, suggests that a \textit{graviton } exchange between two
quantum superposed masses can provide a bonafide quantum feature if
the two quantum systems are allowed to interact \textit{solely} gravitationally.
In fact, the two quantum systems can be entangled by a graviton exchange in a Feynman diagram, which can be tested via a scheme 
known as the quantum gravity induced entanglement of masses (QGEM)~\cite{Bose:2017nin,Marshman:2019sne}.
In any given experiment, the
number of gravitons, either on-shell or off-shell, can leave a detectable mark on observations. In most typical cases, for general open systems, the number of quanta is also indicative of whether a quantum system can be approximated by the mean field or a statistical ensemble~\footnote{Number or density plays an important role, there are roughly $400$
microwave photons per cubic centimetre, and its fluctuations have
been discovered in terms of two-point temperature correlations by
the latest Planck experiments~\cite{Aghanim:2018eyx}. However, we
still do not know whether these fluctuations are classical or quantum~\cite{Martin:2017zxs,Ashoorioon:2012kh}.}.

This paper aims to point out that there is a universal bound in any
quantum interaction of a graviton with the matter degrees of freedom.
For the time being, we neglect any other interaction besides gravity.
This universal behaviour can be studied by \textit{tracing out} the
matter degrees of freedom -- we will show that the occupation number
of gravitons is always proportional to the Area-law of such a gravitational
system. Intriguingly,
such a behaviour can be thought to be universal from the point of
view of Bekenstein's bound on a gravitational entropy~\cite{Bekenstein:1980jp}
(while it has been argued that physically realistic local quantum Hamiltonians in the ground state
follow the Area-law of entanglement entropy~\cite{Bombelli:1986rw,Srednicki:1993im}).
Since in a thermodynamic system, the entropy is proportional to the number of gravitons, and
entropy is saturated by the Area-law, no wonder by tracing out all
the matter degrees of freedom, we obtain that the Area-law always
bounds the occupation number of gravitons.

%%%%%%%%%%%%%%%%%%%%%

\textbf{Graviton coupling to matter:} To set up our computation, let
us assume that we work in a world line of a particle in a Fermi-normal
coordinate, and without loss of generality, let us assume that the
particle motion is in the $x$-axis. Let us consider an ideal matter
following a geodesic trajectory, $x^{\mu}$ (the situation of an observer
following a generic time-like curve can be analysed similarly without
affecting the final results). We will start from the general relativistic
point-particle Lagrangian\footnote{We assume $c=1$ here, and we work with $(-,+,+,+)$ signature,and
$\mu=0,1,2,3$.}: 
\begin{equation}
L=-m\sqrt{g_{\mu\nu}\dot{x}^{\mu}\dot{x}^{\nu}},\label{eq:lagrangian}
\end{equation}
where $m$ is the mass of the system, and $g_{\mu\nu}$ is the metric
expressed in Fermi normal coordinates. In particular, we will write
the metric as $g_{\mu\nu}=\eta_{\mu\nu}+h_{\mu\nu},$ $\eta_{\mu\nu}$
is the Minkowski metric, and $h_{\mu\nu}$ is the spacetime curvature
perturbation near the geodesic up to order $\mathcal{O}(x^{2})$.
%~\cite{visser2018post}.

Let us assume that the matter is moving slowly, the dominant contribution
to the dynamics will be given by~\cite{Rakhmanov:2014noa}: 
\begin{equation}
g_{00}=-(1+\ddot{h}_{11}x^{2}),\label{eq:gtt}
\end{equation}
where $\ddot{h}_{11}=2R_{0101}$ is the ``+'' component of the gravitational
waves usually discussed in the transverse-traceless (TT) coordinates,
and $R$ is the Riemann tensor (here $\ddot{h}_{11}\equiv\ddot{h}_{11}(t,0)$
denotes only a number evaluated on the reference geodesic).

Now, let is use Eqs.~(\ref{eq:lagrangian}) and (\ref{eq:gtt}),
we can then readily find the interaction Lagrangian between graviton
and matter degrees of freedom: 
\begin{equation}
L_{\text{int}}=\frac{m}{4}\ddot{h}_{11}x^{2}.\label{eq:intlagrangian}
\end{equation}
Let us now expand the gravitational fluctuation in terms of Fourier
modes\footnote{The notion of Fourier modes implicitly assumes that we are in an asymptotically flat
space times, and free from curvature singularities.}, see Refs.~\cite{Weinberg1972}: 
\begin{equation}
{h}_{ij}(t,\bm{x})=\int d\bm{k} \sqrt{\frac{G\hbar}{\pi^{2}\omega_{k}}} {g}_{\bm{k},\lambda}\mathtt{e}_{ij}^{\lambda}(\bm{n})e^{-i(\omega_{k}t-\bm{k}\cdot\bm{x})}+\text{H.c.},\label{eq:gravfield}
\end{equation}
where $G$ is the Newton's constant, $\omega_{k}=k$, $k=\|\bm{k}\|$,
$\bm{n}=\bm{k}/\|\bm{k}\|$, and ${g}_{\bm{k},\lambda}$ is the annihilation
operator. In Eq.~\eqref{eq:gravfield} we also implicitly assume
the summation over the polarizations\footnote{\label{f4}The basis tensors satisfy the completeness relation: $\sum_{\lambda}\mathtt{e}_{ij}^{\lambda}(\bm{n})\mathtt{e}_{kl}^{\lambda}(\bm{n})=P_{ik}P_{jl}+P_{il}P_{jk}-P_{ij}P_{kl}$
where $P_{ij}\equiv P_{ij}(\bm{n})=\delta_{ij}-\bm{n}_{i}\bm{n}_{j}$.
For later convenience we write the integral: $\int d\bm{n}\,P_{11}(\bm{n})P_{11}(\bm{n})=32\pi/15$.}, $\sum_{\lambda}$, where $\mathtt{e}_{jk}^{\lambda}$
denote the basis tensors for the two polarizations,$\lambda=1,2$. From Eq.~\eqref{eq:intlagrangian} and \eqref{eq:gravfield} we
however see that only $\mathtt{e}_{11}^{\lambda}(\bm{n})$ is relevant
and we can write also the corresponding kinetic term for the mass-less
graviton field to be: 
\begin{eqnarray}
H_{\text{grav}} & = & \int d\bm{k}\,\hbar\omega_{k}g_{\bm{k},\lambda}^{\dagger}{g}_{\bm{k},\lambda}\nonumber \\
 & = & \sum_{\lambda}\int d\bm{k}\frac{\hbar\omega_{k}}{4}\left[{P}_{\bm{k},\lambda}^{2}+{Y}_{\bm{k},\lambda}^{2}\right],\label{eq:Hgrav}
\end{eqnarray}
where 
\begin{equation}
{Y}_{\bm{k},\lambda}={g}_{\bm{k},\lambda}+{g}_{\bm{k},\lambda}^{\dagger}\qquad{P}_{\bm{k},\lambda}=i({g}_{\bm{k},\lambda}^{\dagger}-{g}_{\bm{k},\lambda}).\label{eq:2}
\end{equation}
The interaction Hamiltonian can be derived from the interaction Lagrangian
in Eq.~(\ref{eq:intlagrangian}), and by using Eq.~(\ref{eq:gravfield}),
we obtain~\cite{Toros:2020dbf2}: 
\begin{equation}
{H}_{\text{int}}=-m\sum_{\lambda}\int d\bm{k}e_{11}^{\lambda}(\bm{n})\mathcal{C}_{\bm{k}}{Y}_{\bm{k},\lambda}{x}^{2},\label{eq:interaction}
\end{equation}
where $x$ is the position operator of the particle, and 
\begin{equation}
\mathcal{C}_{\bm{k}}^{\lambda}=\sqrt{\frac{G\hbar\omega_{k}^{3}}{16\pi^{2}}}.\label{eq:cklambda}
\end{equation}
The interaction in Eq.~(\ref{eq:interaction}) assumes long-wavelength
gravitons such that the cutoff frequency $\bar{\omega}=2\pi/l$.  We
note that by combining Eqs.~(\ref{eq:Hgrav}) and (\ref{eq:interaction}),
we find the relevant part of the gravitational Hamiltonian: 
\begin{eqnarray}\label{quant-int}
H & = & \sum_{\lambda}\int d\bm{k}\frac{\hbar\omega_{k}}{4}\left[{P}_{\bm{k},\lambda}^{2}+{Y}_{\bm{k},\lambda}^{2}\right]\nonumber \\
 &  & -m\sum_{\lambda}\int d\bm{k}e_{11}^{\lambda}(\bm{n})\mathcal{C}_{\bm{k}}{Y}_{\bm{k},\lambda}{x}^{2}\,.
\end{eqnarray}
%Here, we note that we are assuming that the matter Hamiltonian is in a ground state.
The key points here are that we treat both matter and graviton 
on an equal footing, when it comes to quantum in nature, and we are dealing with a self-gravitating 
quantum system of mass $m$, whose interaction with graviton is determined by Eqs.~(\ref{eq:interaction}).
%Besides this, we will also have the matter Hamiltonian, which we consider here to be a simple Harmonic oscillator without losing any generality.
%\begin{equation}
%H_{m}=\frac{p^{2}}{2m}+\frac{m\omega_{m}^{2}}{2}x^{2}\,.
%\end{equation}
%We have now obtained a complete Hamiltonian which will describe both gravity and matter, and we intend to treat both of them on an equal footing.

%%%%%%%%%%%%%%%%%%%%%%%%%%%%%%%%%%%

\textbf{Tracing out matter degrees of freedom:} Suppose we consider
the mean field approximation \emph{of the matter sector} in Eq.~(\ref{quant-int}),
i.e. ${x}^{2}\rightarrow\langle{x}^{2}\rangle$, which is applicable in our case, because the 
gravitational coupling to matter degrees of freedom is very weak, suppressed by $G\sim 1/M_p^2$. Furthermore, we are 
far away from any pathologies in space time, which allows us to work within the leading order contribution in 
$G\sim 1/M_p^2$, it is a good assumption that the states of the graviton and the matter sector are factorizable.
From Eq.~(\ref{quant-int}) we thus find a displaced harmonic trap for the gravitational field:
\begin{equation}
{H}=\sum_{\lambda}\int d\bm{k}\frac{\hbar\omega_{k}}{4}\Biggl[{P}_{\bm{k},\lambda}^{2}+({Y}_{\bm{k},\lambda}-\alpha_{\lambda,\bm{k}})^{2}\Biggr],\label{eq:total2}
\end{equation}
where the center is given by
$\alpha_{\lambda,\bm{k}}\equiv\frac{2m}{\hbar\omega_{k}}e_{11}^{\lambda}(\bm{n})\mathcal{C}_{\bm{k}}\langle{x}^{2}\rangle$. 
Now, let us suppose that the gravitational field mode ${g}_{\bm{k},\lambda}$
is in a ground state, centred around $\alpha_{\bm{k}}^{\lambda}$,
which can be described by a displaced coherent state,
\begin{equation}
\vert\alpha_{\bm{k}}^{\lambda}\rangle=D(\alpha_{\bm{k}}^{\lambda})\vert0\rangle=e^{\alpha_{\bm{k}}^{\lambda}\left[{g}_{\bm{k},\lambda}^{\dagger}+{g}_{\bm{k},\lambda}\right]}\vert0\rangle.
\end{equation}
By choosing the gravitational field to be in the ground state of the displaced harmonic trap we envisage that the matter system and the gravitational field have reached a steady-state --  a different choice for the gravitational field state will not change significantly the final result as long as the state remains centered and confined around the same minimum.
For such a displaced quantum state we can compute the expectation values  $\langle~\cdot~\rangle $ --  here we will be interested in the mean value and fluctuations of the number operator $N_{\bm{k},\lambda}\equiv {g}_{\bm{k},\lambda}^{\dagger}{g}_{\bm{k},\lambda} $, where the gravitons follow the commutation relations: $[g_{\bm{k},\lambda},g^\dagger_{\bm{k}',\lambda'}]=\delta(\bm{k}-\bm{k}')\delta_{\lambda,\lambda'}$. 

We find the occupation number of ${g}_{\bm{k},\lambda}$ gravitons in the ground state to be:
\begin{equation}
\langle N_{\bm{k},\lambda}\rangle=\vert\alpha_{\bm{k}}^{\lambda}\vert^{2}=\left(\frac{2m}{\hbar\omega_{k}}e_{11}^{\lambda}(\bm{n})\mathcal{C}_{\bm{k}}\langle{x}^{2}\rangle\right)^{2}\,.
\end{equation}
The total number of gravitons can be then computed by summing and integrating over all the  graviton modes: 
\begin{equation}
N_{g}\equiv  \sum_{\lambda}\int d\bm{k} \langle N_{\bm{k},\lambda} \rangle=\sum_{\lambda}\int d\bm{k}\left[\frac{2m}{\hbar\omega_{k}}e_{11}^{\lambda}(\bm{n})\mathcal{C}_{\bm{k}}\langle{x}^{2}\rangle\right]^{2}.\label{eq:totalgravitonnumber}
\end{equation}
By using Eq.~(\ref{eq:cklambda}) and footnote \ref{f4}, we note that 
\begin{equation}
N_{g}=\int {k^{2}dk}\frac{4m^{2}}{\hbar^{2}\omega_{k}^{2}}{\int d\bm{n}\sum_{\lambda}e_{11}^{\lambda}(\bm{n})e_{11}^{\lambda}(\bm{n})}{\mathcal{C}_{\bm{k}}^{2}}\langle{x}^{2}\rangle^{2}\,,
%N_{g}=\int\underbrace{k^{2}dk}_{\frac{\omega_{k}^{2}}{c^{3}}d\omega_{k}}\,\frac{4m^{2}}{\hbar^{2}\omega_{k}^{2}}\underbrace{\int d\bm{n}\sum_{\lambda}e_{11}^{\lambda}(\bm{n})e_{11}^{\lambda}(\bm{n})}_{\frac{32\pi}{15}}\underbrace{\mathcal{C}_{\bm{k}}^{2}}_{\frac{G\hbar\omega_{k}^{3}}{16\pi^{2}c^{2}}}\langle{x}^{2}\rangle^{2}\,,
\label{eq:totalgravitonnumber1}
\end{equation}
reduces to 
\begin{equation}
N_{g}=\frac{8m^{2}G}{15\pi\hbar}\langle{x}^{2}\rangle^{2}\int_{0}^{\bar{\omega}}d\omega_{k}\omega_{k}^{3},\label{eq:totalgravitonnumber2}
\end{equation}
where $\bar{\omega}$ is the cutoff frequency for a box of side $l$.

%%%%%%%%%%%%%%%%%%%%%%%%%%%%%%%%%%%%%%%%%%%%%%%%%

\textbf{Number of gravitons:} We can perform the integration
in Eq.~\eqref{eq:totalgravitonnumber2}, and by using $\bar{\omega}=2\pi/l$,
we find a factor $(2\pi/l)^{4}/4$. Furthermore, we note that $\langle{x}^{2}\rangle^{2}\leq l^{4}/2^{4}$.
We then find that the $l$ dependence goes away from the
numerator and the denominator in Eq.~(\ref{eq:totalgravitonnumber2}), leaving us with a straightforward relationship,
where the number of gravitons is simply given by the mass $m$, and
Newton's constant, $G$:
\begin{equation}
N_{g}\leq \frac{4\pi^{3}}{30}\frac{Gm^{2}}{\hbar}=\frac{4\pi^{2}}{30}\left(\frac{\text{Area}}{4G\hbar}\right)\,.\label{eq:last}
\end{equation}
In the above, by area we mean the area corresponding to the Schwarzschild radius
of mass $m$, given by $R_{s}=2Gm$, which is also the natural length-scale over which the above number of gravitons are confined. 
Same result, aside minor numerical factors, is obtained for a relativistic energy-momentum tensor.~\footnote{Suppose we had taken an example of a relativistic energy-momentum tensor $T_{\mu\nu}$, where the graviton-matter interaction would be dictated by the interaction $\sqrt{G}h_{\mu\nu}T^{\mu\nu}$, and we would then quantise the graviton in a transverse traceless (TT)-gauge,  we would get a similar answer barring the numerical factors, $N_g\leq (32\pi^3/105)(Gm^2/\hbar)$. }

Since the entropy of a black
hole is proportional to the Area of the hole, the number of gravitons
found above naturally scales as Bekenstein's entropy $N_{g}\sim S_{g}$, or specifically,
it is consistent with the bound set by Bekenstein, i.e. $N_{g}\sim S_{g}<2\pi ER/\hbar$,
where $E$ is total mass-energy of the system, and $R$ is the characteristic
radius, in the black hole case $ER\sim2Gm^{2}$~\cite{Bekenstein:1980jp}. 
 At this point, it has not avoided our attention that the gravitational entropy is indeed holographic~\cite{tHooft:1993dmi,Susskind:1994vu}.

An intriguing feature of the above expression is that the number of gravitons signifies how
we may be able to quantify the quantum nature of space-time itself.
If we can squeeze the matter within the Schwarzschild radius, the
number of gravitons saturate the black hole entropy. 
In GR, Schwarzschild radius is the only length scale that appears if we place a mass $m$ in an asymptotically flat space-time.
In fact, from Eq.~\eqref{eq:last} we
can roughly estimate the number of gravitons to be: 
\begin{equation}
N_{g}\approx\left({m}/{M_{p}}\right)^{2}\,.\label{simi}
\end{equation}
For a solar mass blackhole $m\sim10^{33}$g, and $M_{p}\sim10^{-5}$g,
$N_{g}\sim10^{76}$. The large $N_{g}\geq1$, sets a barrier for
any massive quantum system, whose mass exceeds that of the gravitational
mass $m\geq M_{p}\sim10^{-5}$g. From Bohr's correspondence principle we may be able to treat such self-gravitating system approaching towards a classical limit.

The notion that large values of $N_g$ can be associated with classical behaviour is further reinforced by computing the quantum fluctuations,
\begin{equation}
\Delta N_{g} \equiv \sqrt{\langle N^2  \rangle  -  \langle N \rangle^2 }, \label{DeltaN}
\end{equation}
of the total number operator $N\equiv \sum_\lambda \int d\bm{k} N_{\bm{k},\lambda}$.
We have already calculated $\langle N\rangle = N_g$ and using the commutation relations $[g_{\bm{k},\lambda},g^\dagger_{\bm{k}',\lambda'}]=\delta(\bm{k}-\bm{k}')\delta_{\lambda,\lambda'}$ we can show $\langle N^2\rangle = N_g + N_g^2$  (see for example \cite{Gerry:2005}). Inserting the expectation values back in Eq.~\eqref{DeltaN} we then find that the fluctuations grow only as $\Delta N_{g}= \sqrt{N_g}$. Hence we find that the relative quantum fluctuations follow as 
\begin{equation}
\Delta N_g/N_g\sim 1/\sqrt{N_g}
\end{equation}
 in line with the expectation that the system appears classical for large values of $N_g\gg 1$. 

A curious reader would wonder if we were to trace the graviton degrees of freedom instead of matter degrees of freedom. In such a case, $Y_{{\bm k},\lambda}\rightarrow \langle Y_{{\bm k},\lambda}\rangle$ in Eq.~\eqref{quant-int}, and we would be left with a simple harmonic oscillator potential. For such a system, the occupation number $N_m$ will not follow the Area-law. We can further ask whether the Area-law is a generic property for all massless fields or specific to the matter-graviton coupling. We recall that the quadratic nature of the coupling $\sim x^2$ in Eq.~\eqref{eq:intlagrangian} was critical: we have shown $N_g \propto \langle x^2 \rangle$ -- where $\langle x^2 \rangle$ is always non-zero -- leading to the Area-law. In contrast, a linear coupling, e.g., $\sim x$, can always be cancelled by a change of the reference frame and should thus not play a role in the derivation of the Area-law. More generally one can expect the emergence of an Area law whenever the coupling gives rise to a length scale, while for other types of interactions (such as the electromagnetic) one is left without such interpretation.

%%%%%%%%%%%%%%%%%%%%

%%%%%%%

\textbf{Applications and features:} 
Let us now provide  another intriguing connection to what we have just found in Eqs.~(\ref{eq:last}) and (\ref{simi}).
 In Refs.~\cite{Dvali:2011aa,Dvali:2014ila,Dvali:2020wqi,Casadio:2021cbv}, see Ref.~\cite{Ruffini} for earlier discussions,
a very interesting idea has been developed, a corpuscular nature of
a black hole. A black hole is perceived to be a Bose-Einstein condensate
of a large number of weakly interacting gravitons, where the authors
have found exactly a similar scaling as in Eq.~(\ref{simi}), but
from a different point of view. They did not have to trace out the
matter degrees of freedom explicitly. Instead, they argued that a
black hole could be justly described by an ensemble of $N_{g}\gg1$
gravitons. The authors further argued that the number of gravitons
will suppress the effective interaction of a gravitons, i.e., given
by $\alpha_{g}\leq1/N_{g}$. For a large $N_{g}\gg1$, the black hole
behaves like a coherent system, albeit a leaky one. The graviton escape
signifies the Hawking evaporation of a black hole. Such a gravitational
system with $N_{g}\gg1$ can be perceived to be approaching a classicalisation
limit, where extracting quantum features will be becoming extremely hard.

The other limiting case, when $N_{g}\ll1$, provides us with an intriguing
possibility of extracting the quantum nature of a graviton. This limit
arises when $m\leq10^{-5}$g, for such a system, the number of gravitons
is less than one irrespective of the size of the box. 
Therefore, it seems that nature provides us with a window of opportunity where it might
be possible to construct experiments cleverly to study the quantum
nature of gravity for $m\leq10^{-5}$g. Indeed, if $m\ll10^{-5}\text{g}$,
it will be again harder to probe or extract any quantum features.
It has been experimentally possible to create macromolecules
$m\sim10^{-19}$~g over spatial superposition for $\sim0.25$~$\mu$m,
or atoms $m\sim10^{-21}$~g over $\sim0.5$~m)~\cite{Kovachy:2015,Arndt:2019}.
In all these cases, $N_{g}\sim10^{-28}$, or $10^{-32}$, way too
small to observe any detectable features of quantum gravity.

The situation appears to be quite familiar in cosmology as well if we were to assume that a world-line can describe the entire observable patch of the Universe. In that case, the mass contained in the Universe would be $m\sim \rho\times V$, where $\rho=H^2(t)M_p^2$ is the constant energy density of the Universe, and $V\sim H^{-3}(t)$ is the observable volume, which would scale with Hubble expansion rate $H(t)$ of the Universe. If we were now to compute the occupation number of gravitons within my observable patch of the Universe, we would find,
\begin{equation}
N_{g}\sim\left({M_{p}}/{H(t)}\right)^{2}\,.
\end{equation}
In the current Universe, $H\sim10^{-42}$~GeV, which gives $N_{g}\sim10^{121}$, which indicates that the space-time 
 is perhaps very close to a classical description, and the number of gravitons again scales as the Area-law, $N_g\sim {\rm Area/4G}$, because 
the observable area of the Universe will scale as $\sim H^{-2}(t)$. 

On the other hand, we already have a constraint on the largest scale
in the Universe indirectly. Since we have not seen any primordial gravitational
waves, or the B-mode polarisation in the cosmic microwave background
radiation, we believe that the scale of cosmic inflation cannot be
arbitrarily large. The latest Planck data places an upper bound on
$H\sim H_{inf}\sim10^{13}$~GeV~\cite{Aghanim:2018eyx}. Of course,
we have invoked here primordial inflation as a mechanism to seed the
fluctuations in the cosmic microwave background radiation. However,
even for such a high energy probe, the number of gravitons present
in the Hubble radius during inflation would behave nearly classically, i.e.
$N_{g}\sim10^{12}$, and $\Delta N_g/N_g\sim 10^{-6}$.

Finally, we can ask how robust our analysis is with regards to the nature of
classical gravity. Note that we have not assumed any specific form of gravitational
action. 
The only assumption we have made here is that the gravitons can be described by the massless degrees of freedom in a harmonic oscillator state with the minimal coupling to matter degrees of freedom given by $G\sim 1/M_p^2$, which is valid for gravitational theories beyond 3 spatial dimensions.
However, this argument will change if we 
insist on studying any higher derivative extension of gravity, which will bring inevitably 
a new scale in the problem, say $M_s < M_p$ in four space time dimensions, see~\cite{Biswas:2011ar,Buoninfante:2019swn}.
For such class of theories of gravity, we will have to tread the 
occupation number of gravitons carefully case by case.

\textbf{Conclusion:} We briefly conclude by highlighting
that we have provided a rather model-independent constraint on the
occupation number of gravitons in a quantum system determined by its
mass and $M_{p}$ by tracing out the matter degrees of freedom. The
occupation number is bounded by the Area-law, which is a reminiscence
to Bekenstein's bound. Our bound suggests that the occupation number
of gravitons in the black hole geometry will be bounded by the Area
of a black hole, which is also the gravitational entropy of the object.
For mass $m\geq10^{-5}$g, the number of gravitons occupied within
the gravitational radius, i.e. the Schwarzschild radius, is much larger
than one. In an optimal bound on
mass $m\sim10{}^{-5}$g would be ideal for extracting any quantum
behaviour of a graviton. Furthermore, tracing out the matter degrees of freedom renders the 
space-time fairly classical within the observable patch of the Universe.

\textit{Acknowledgements:} Authors would like to thank Roberto Casadio for helpful discussions. SB would like to acknowledge EPSRC grants No. EP/N031105/1 and EP/S000267/1.
AM's research is funded by the Netherlands Organisation for Science and Research (NWO) grant number
680-91-119. MT acknowledges funding by the Leverhulme Trust (RPG-2020-197).

%%%%%%%%%%%%%%%%%%%%%%%%%%%%%%

\end{document}